\documentstyle[12pt,a4wide]{article}
\title{\mbox{\LARGE{Friction Domination with Superconducting Strings}}}
\author{
\mbox{\large{Konstantinos Dimopoulos}}
\thanks{e-mail: K.Dimopoulos@damtp.cam.ac.uk}
\vspace{0.2cm}\\
\mbox{\large{and}}
\vspace{0.2cm}\\
\mbox{\large{Anne--Christine Davis}}
\thanks{e-mail: A.C.Davis@damtp.cam.ac.uk}
\vspace{0.4cm}\\
\mbox{\normalsize{\em Department of Applied Mathematics and
Theoretical Physics,}}\\
\mbox{\normalsize{\em University of Cambridge, Silver Street,}}\\
\mbox{\normalsize{\em Cambridge, CB3 9EW, U.K.}}
}

\begin{document}
\begin{titlepage}
\maketitle
\begin{abstract}
We investigate the evolution of a superconducting string network with
arbitrary, constant string current in the friction dominated
regime. In the absence of an external magnetic field the network
always reaches a scaling solution. However, for string current
stronger than a critical value, it is different than the usual,
horizon--scaling of the non-superconducting string case. In this case
the friction domination era never ends. Whilst the superconducting
string network can be much denser than usually assumed, it can never
dominate the universe energy density. It can, however, influence the
cosmic microwave background radiation and the formation of large scale
structure. When embedded in a primordial magnetic field of sufficient
strength, the network never reaches scaling and, thus, eventually
dominates the universe evolution. 
\end{abstract}
\vspace{1cm}
\flushright{DAMTP-97-8}
\end{titlepage}

{\bf 1. Introduction}

\bigskip

The microphysics of cosmic strings has received considerable attention.
In particular, Witten \cite{witt} showed that cosmic strings become
superconducting as a result of boson condensates or fermion zero modes
in the string core. Such strings are capable of carrying a sizeable 
current, with the maximum current being about  \mbox{$10^{20}A$}
for a grand unified scale string. Inevitably, such currents have
cosmological and astrophysical \cite{astro} consequences. The
consequences for emission of synchrotron radiation  \cite{field} and
for high energy $\gamma$-rays \cite{schram}\cite{ngrays}\cite{grays}
have been explored. 

Unlike non-conducting strings, loops of superconducting string can
be stabilised from collapse by the angular momentum of the current
carriers, forming vortons \cite{vortons}. If vortons are sufficiently
stable then consistency with standard cosmology puts severe constraints on 
theories giving rise to such strings. However,
such constraints have been derived assuming that the evolution of a
network of superconducting strings is similar to that of ordinary
strings. Early studies using both analytic \cite{evol} and numerical
techniques \cite{amst1}\cite{matz}\cite{nat2}
showed that the string evolution was indeed similar
to that of ordinary cosmic strings. However, these studies neglected 
the very early times when the string is interacting strongly with the
surrounding plasma. During this friction dominated period the string
correlation length grows until is catches up with the horizon and a
scaling solution sets in.  

Since the interaction of particles with a superconducting string
\cite{field}\cite{warren}
is very different from that of an ordinary string \cite{ever},
there is every reason to expect that the friction dominated period
could be vastly different. Thus, it is important to investigate
this and ascertain whether or not the superconducting string network
really does reach a scaling solution. Even if the network does reach
a scaling solution it could still be very different from that of the
non-conducting case. 

We first discuss the interaction of the plasma particles on the string
and then show the evolution of the curvature radius assuming a constant
string current. We show that, in 
the absence of a primordial magnetic field, there is a critical current
above which the friction dominated period never ends. Instead the string
reaches a so-called plasma scaling solution, where the density of strings
is considerably greater than the usual horizon scaling. The cosmological
consequences of this are investigated. When the string is embedded in
a magnetic field of strength greater than a critical value, a string
dominated universe results. Finally, we discuss our assumption that the
string current is constant. Our conclusions are summarised in the final
section. In what follows, unless stated otherwise, we use
natural units (\mbox{$\hbar=c=1$}).

\bigskip

{\bf 2. Friction on superconducting cosmic strings}

\nopagebreak[4]

\bigskip

\nopagebreak[4]

Friction on a cosmic string is caused by particle interaction
with the string as it moves through the
plasma. The friction force per unit length is,

\begin{equation}
f\sim \rho\sigma v \bar{v}
\label{f}
\end{equation}
where $\rho$ is the energy density of the plasma, 
$\sigma$ is the interaction cross-section,
$v$ is the velocity
of the string segment and \mbox{$\bar{v}\simeq\Delta p/m$} is given by,

\begin{equation}
\bar{v}\equiv \max \{ v,v_{th}\}
\label{vbar}
\end{equation}
where $\Delta p$ is the particle's momentum change,
\mbox{$v_{th}\sim\sqrt{T/m}$}  is the thermal velocity of the plasma
particles,  \mbox{$m^{2}\simeq m_{0}^{2}+\alpha
T^{2}$} is the mass of a  plasma particle with rest mass $m_{0}$,
(\mbox{$\alpha = g^{2}/4\pi$} with $g$ the gauge
coupling) and $T$ is the plasma temperature. Also, \mbox{$m_{P}=
1.22\times 10^{19}GeV$} is the Planck mass.

Curves and wiggles on the strings tend to untangle due to string
tension, which results in oscillations of the curved sting segments
on scales smaller than the causal horizon and larger than their
curvature radii. Friction dissipates the energy of these oscillations
and leads to their gradual damping. Thus, the stings become smooth
on larger and larger scales with their curvature radius growing
accordingly. The characteristic damping timescale is \cite{kibb},

\begin{equation}
t_{d}\equiv\frac{\mu v^{2}}{\dot{\cal E}}\sim\frac{\mu
v}{f}\sim\frac{\mu}{\rho\sigma\bar{v}} 
\label{td}
\end{equation}
where $\mu$ is the string mass per unit length and
\mbox{$\dot{\cal E}\sim fv$}, is the energy loss per unit time
per unit length. 
The above timescale corresponds to a length-scale
called the friction length \cite{hin}\cite{carl}.

Initially the friction length is smaller than the curvature radius.
In this case Kibble \cite{kibb} has shown that the growth of the
string curvature radius is,

\begin{equation}
\frac{dR}{dt}\sim\frac{t_{d}}{R}
\label{R}
\end{equation}

If the growth of the friction length is faster than that of $R$,
then at some point it will catch up with the curvature radius and
(\ref{R}) will cease to be valid. This
occurs when the curvature radius reaches the horizon \cite{kibb} 
and the strings become smooth on horizon scales. From then on
\mbox{$R\sim t$}, the friction domination era
ends and the string network satisfies the well--known horizon--scaling
solution. Thus, in order to see whether a network of superconducting
cosmic strings ever reaches a scaling solution one has to study the
evolution of the string curvature radius. For this, a closer look to
the superconducting string system is required.

The current of a charged--current carrying superconducting string
creates a magnetic field around the string core. This field is very
strong near the string and does not allow the plasma particles to
approach. As the charged particles encounter the magnetic field their
motion is diverted in such a way that they create a surface current of
opposite orientation than the string current \cite{field}, which
screens the magnetic field. Thus, while the string moves in the plasma
it is shielded by a magnetocylinder, which contains the
magnetic field and does not allow the plasma to penetrate.

The flow of the plasma around the string creates a shock front, which
is the border of the magnetocylinder \cite{field}, 
whose distance from the string is determined by the pressure balance
between the magnetic field and the plasma. The string magnetic field
is given by the Biot-Savart law,\footnote{For temperatures higher than
the electroweak 
energy scale the electroweak symmetry is unbroken and the ``magnetic''
field of the string is actually due to the hypercharge
generator. However, as shown in \cite{EW}, its magnitude still
follows, \mbox{$B\sim J/r$} as in (\ref{Bs}).}

\begin{equation}
B_{s}(r)\simeq\frac{2J}{r}
\label{Bs}
\end{equation}
where $J$ is the string current and $r$ is the distance from the
string. All through this paper, unless stated otherwise, we will
assume that the string current $J$ is constant. The validity of
this assumption will be discussed later but here we can just mention
that it is primarily based on conservation laws due to topological
index theorems \cite{witt}. The magnitude of the current will be
treated as a free parameter and can have any value up to
\mbox{$J_{max}\sim e\sqrt{\mu}$}, where $e$ is the charge of
the current carriers. We also assume, for simplicity, that the
current switches on at the phase transition that creates the string
network.

The pressure balance, \mbox{$B_{s}^{2}(r_{s})\sim\rho\bar{v}^{2}$}
suggests that the dimensions of the magnetocylinder are of order,

\begin{equation}
r_{s}\sim\frac{J}{\sqrt{\rho}\,\bar{v}}
\label{rs}
\end{equation}

The magnetocylinder cross-section is not circular of course
\cite{hurley} but (\ref{rs}) is a good estimate of the minimum
distance of the shock front from the string \cite{field}. We can use,
therefore, $r_{s}$ as the superconducting, charged-current carrying
string cross-section. From (\ref{rs}) it is evident that the
cross-section increases with the string current, in agreement with
\cite{warren}. 

The above cross-section $r_{s}$ has to be compared
with the usual cross-section for a cosmic string,

\begin{equation}
\sigma_{cs}\sim k^{-1}
\label{sns}
\end{equation}
where \mbox{$k\sim(m\bar{v})^{-1}$} is the momentum of the incident
particle in the frame of the string \cite{ever}.

\bigskip

{\bf 3. Evolution of the curvature radius}

\nopagebreak[4]

\bigskip

\nopagebreak[4]

The evolution of the string curvature radius determines whether the
string network reaches a scaling solution or overcloses the universe
instead.  

As suggested by numerical simulations \cite{matz}, the intercommuting
process of superconducting strings is very similar to the
non-superconducting string case, i.e. the probability of producing
string loops is of order unity. High intercommutation probability
ensures the scaling of the network in the absence of friction. 
Moreover, since the superconducting network dissipates its energy more
efficiently, due to additional radiation emission \cite{evol}, the
intercommutation probability can be much lower 
than the limit required for non-superconducting strings without
undermining the scaling solution. 

Assuming an intercommutation probability of order unity suggests that
the network density evolves according to, 

\begin{equation}
\frac{\dot{\rho}_{s}}{\rho_{s}}=-2\,\frac{\dot{a}}{a}-\frac{v}{R}
\label{above}
\end{equation}
where $a$ is the scale factor of the universe (\mbox{$a\propto
t^{1/2}$} for the radiation era).

The above implies that the typical inter-string distance grows as
{\mbox{$\dot{R}\sim v$}. From (\ref{above}) it can be inferred that
the network would satisfy a
scaling solution, \mbox{$\rho_{s}/\rho = const.$}, provided
\mbox{$R\propto t$} for the radiation era. If $R$ manages to grow up
to horizon size then scaling is ensured since \mbox{$R\sim t$}. 
In this case the inter-string distance is the horizon size and the
string velocity is \mbox{$v\sim 1$}. We will refer to this scaling
solution as horizon--scaling.

From (\ref{f}) and (\ref{sns}) we find that the friction force 
for non-superconducting strings is,

\begin{equation}
f_{ns}\sim\frac{\rho v}{m}
\label{fns}
\end{equation}

Similarly, from (\ref{f}) and (\ref{rs}) we obtain for the plasma
friction force,

\begin{equation}
f_{pl}\sim Jv\sqrt{\rho}
\label{fpl}
\end{equation}

Note that the friction forces
are not affected by the transition from the radiation to the matter
era, since \mbox{$\rho\sim (m_{P}/t)^{2}$} at all times.

Inserting the above into (\ref{td}) one finds,

\begin{equation}
t_{d}^{ns}\sim\frac{\mu m}{\rho}
\label{tdns}
\end{equation}

and

\begin{equation}
t_{d}^{pl}\sim\frac{\mu}{J\sqrt{\rho}}
\label{tdpl}
\end{equation}

From the above it can be inferred that the temperature dependence of
the damping time is changed only due to the variation of the particle
mass. This has an effect only when the plasma friction force is
subdominant. In this case, using (\ref{R}) we obtain,

\begin{equation}
R_{ns}\sim\frac{\eta}{m_{P}^{3/4}}\,t^{5/4}
\label{Rns1}
\end{equation}

and

\begin{equation}
R'_{ns}\sim [\frac{\eta}{\sqrt{m_{0}\,m_{P}}}]\,t_{m}^{-1/2}t^{3/2}
\label{Rns2}
\end{equation}
where (\ref{Rns1}) and (\ref{Rns2}) correspond to temperatures higher
and lower than \mbox{$T(t_{m})\sim m_{0}$} respectively\footnote{The
prime when used denotes the low temperature case.}
(\mbox{$m_{0}\sim 1\,GeV$}) and we have used that,
\mbox{$\mu\sim\eta^{2}$}  with $\eta$ being the scale of the symmetry
breaking that produced the string network. 

Similarly, in the case of plasma friction domination, for all
temperatures,  we have,

\begin{equation}
R_{pl}\sim\frac{\eta}{\sqrt{J\,m_{P}}}\,t
\label{Rpl}
\end{equation}

The above suggest that, if the plasma friction is subdominant, the
curvature radius grows more rapidly than the horizon and
will reach the horizon size at,

\begin{equation}
t_{*}\sim\frac{m_{P}^{3}}{\eta^{4}}
\label{t*}
\end{equation}

or at,

\begin{equation}
t'_{*}\sim\frac{m_{P}^{2}}{\eta^{2}m_{0}}\sim
[\frac{\eta}{\sqrt{m_{0}\,m_{P}}}]^{2}\,t_{*}
\label{t*'}
\end{equation}
for high and low temperatures respectively. Comparing with $t_{m}$ we
find, 

\begin{equation}
\eta\stackrel{>}{\scriptstyle <}\sqrt{m_{0}\,m_{P}}\;\Leftrightarrow\;
t_{*}\stackrel{<}{\scriptstyle >} t'_{*}
\stackrel{<}{\scriptstyle >} t_{m}
\label{ttt}
\end{equation}

Thus, for \mbox{$\eta>\sqrt{m_{0}\,m_{P}}\sim 10^{9}GeV$} the network
would reach horizon scaling before $t_{m}$.

If, on the other hand, plasma friction dominates, then $R$ follows
(\ref{Rpl}) and the curvature radius will always remain in constant
proportion to the horizon size. Also, the strings of the network assume
a constant, terminal velocity of order,

\begin{equation}
v_{T}\sim\dot{R}\sim\frac{\eta}{\sqrt{J\,m_{P}}}
\label{vT}
\end{equation}

Both the friction forces (\ref{fns}) and (\ref{fpl}) are decreasing
with time. However, the plasma friction force declines less rapidly, 
for all temperatures. Thus, there is always a time when
plasma friction will come to dominate. By comparing the forces, for
high and low temperatures respectively, we find,

\begin{equation}
t_{c}\sim\frac{m_{P}}{J^{2}}
\label{tc}
\end{equation}

and

\begin{equation}
t'_{c}\sim\frac{m_{P}}{m_{0}\,J}\sim (\frac{J}{m_{0}})\,t_{c}
\label{tc'}
\end{equation}

However, the evolution of the network will be affected by this only if
\mbox{$t_{c}\leq t_{*}$} (or \mbox{$t'_{c}\leq t'_{*}$}). 
In the opposite case $f_{ns}$ remains
dominant until the curvature radius grows up to horizon size and the
horizon--scaling solution begins. Comparing the two critical times
gives the critical string current,

\begin{equation}
J_{c}\equiv\frac{\eta^{2}}{m_{P}}\sim\eta\sqrt{G\mu}
\label{Jc}
\end{equation}
for all temperatures,  
where $G$ is Newton's gravity constant (\mbox{$G=m_{P}^{-2}$}). 
Note that, for high temperatures,
\mbox{$T_{c}\equiv T(t_{c})\sim J$} and 
\mbox{$J_{c}\sim T_{*}\equiv T(t_{*})$}. 

If the string current is smaller
than $J_{c}$ then the evolution of the curvature radius follows the
usual pattern described in the literature \cite{book}. If, however,
\mbox{$J>J_{c}$}, the curvature radius grows more rapidly than the
horizon until $t_{c}$ (or $t'_{c}$). After that, it follows
(\ref{Rpl}). Since, \mbox{$R_{pl}\propto t$}, the network evolves
again in a self similar  way but with typical inter-string distance
smaller than the usual horizon--scaling by a factor of the magnitude
of the terminal velocity. We will call this scaling solution {\em
plasma--scaling}. 

From the above we also find,

\begin{equation}
J\stackrel{>}{\scriptstyle <} m_{0}\;\Leftrightarrow\;
t_{c}\stackrel{<}{\scriptstyle >} t'_{c}
\stackrel{<}{\scriptstyle >} t_{m}
\label{ttt'}
\end{equation}

Thus, if \mbox{$J>J_{c}$} and \mbox{$J>m_{0}$} the network reaches
plasma--scaling before $t_{m}$. 

Summing up, the above suggest that: 

\begin{itemize}

\item
For \mbox{$J<J_{c}$}, {\em the network always reaches horizon--scaling}
at \mbox{$t=\min\{ t_{*},t'_{*}\}$}.

\item

For \mbox{$J\geq J_{c}$},
{\em the network always reaches plasma--scaling} at 
\mbox{$t=\min\{ t_{c},t'_{c}\}$}.

\end{itemize}

At this point we should point out that,
for currents weaker than \mbox{$J\sim m_{0}\sim 1\,GeV$}, it can be
shown that the magnetocylinder system is not
impenetratable to  plasma particles and, thus, for
\mbox{$J_{c}<J<m_{0}$}, the above analysis is
not entirely reliable.\footnote{From (\ref{Jc}) it follows that
this regime can only be realized
when, \mbox{$\eta\leq\sqrt{m_{0}\,m_{P}}$}.} 

\bigskip

{\bf 4. The plasma--scaling solution}

\nopagebreak[4]

\bigskip

\nopagebreak[4]

Let us, now, estimate the string network's energy density during
plasma--scaling. In the case of horizon--scaling (\mbox{$J\leq
J_{c}$}) the energy density of the network of open strings can be
easily found to be,  

\begin{equation}
\frac{\rho_{s}}{\rho}\sim (\frac{\eta}{m_{P}})^{2}\sim G\mu
\end{equation}

The above result is the usual estimate for a non-superconducting
string network energy density \cite{book}.

Now, if friction continues to dominate (\mbox{$J>J_{c}$}) then
the network reaches plasma--scaling and the energy
density of the open string network is larger. Indeed, using
\mbox{$\rho_{s}\sim\mu/R_{pl}^{2}$} 
we find, 

\begin{equation}
\frac{\rho_{s}}{\rho}\sim\frac{J}{m_{P}}\ll 1
\end{equation}

The above shows that, {\em the current carrying string network can
never dominate the universe energy density}. For overcritical currents
the network density and the string terminal velocity are,

\begin{eqnarray}
J_{c} & \leq\;\; J\;\;\leq & J_{max}\nonumber\\
 & & \nonumber\\
1 & \geq\;\; v_{T}\;\;\geq & \sqrt{\frac{\eta}{m_{P}}}\label{evol}\\
 & & \nonumber\\
G\mu\sim (\frac{\eta}{m_{P}})^{2} & \leq\;
\rho_{s}/\rho
\;\leq & \frac{\eta}{m_{P}}\sim\sqrt{G\mu}\nonumber
\end{eqnarray}

From the above it can be inferred that the larger the current the
slower the strings move. Thus, the inter-string distance and the
curvature radius are smaller. As a result, although
the number of intercommutations per unit time per unit volume
\mbox{$\sim v/R$} \cite{vach} is not reduced, the loops produced  by
the network are smaller and, therefore, less string length
is lost. This results into a larger open string energy
density. However, even the maximum $\rho_{s}$ cannot dominate the 
overall energy density of the universe. 

Still, the existence of an overdense string network could have other
observable consequences. Indeed, were such a network to seed the
large scale structure, it would produce structure of a smaller
correlation. However, it would {\em not} create density inhomogeneities
that may be incompatible with the galaxy formation scenaria. This can be
seen as follows.

Setting \mbox{$\delta\rho\equiv\rho_{s}$}, the fractional density
fluctuation of the string network at horizon crossing is, 

\begin{equation}
(\frac{\delta\rho}{\rho})_{_{S}}\sim\frac{G\mu}{v_{T}^{2}}\geq G\mu
\end{equation}

However, the overdensities generated by the strings are
smaller. Indeed, while moving through the plasma, the open strings
generate wakes of overdense matter due to the conical string metric
\cite{wakes}. The overdensity inside a newly formed wake is of the
order of \mbox{$\delta\rho_{W}/\rho\simeq 1$}, its length is
\mbox{$l_{W}\sim v_{T}\,t$} and its thickness is \mbox{$d_{W}\simeq
v_{T}\,t\tan (\Delta/2)\sim (G\mu)v_{T}\,t$}, where
\mbox{$\Delta\simeq 8\pi G\mu$} is the deficit angle of the string
metric. The linear mass overdensity of a wake is
\mbox{$\delta\mu_{W}=
\delta\rho_{W}\,d_{W}\,l_{W}\sim\rho(G\mu)(v_{T}\,t)^{2}$}.
Thus, since \mbox{$R_{pl}\sim v_{T}\,t$}, 
the total overdensity due to open string wakes is, 

\begin{equation}
(\frac{\delta\rho}{\rho})_{_{W}}\simeq
\frac{1}{\rho}\frac{\delta\mu_{W}}{R_{pl}^{2}}\sim
G\mu
\end{equation}

The above shows that the overdensities generated by the string network
are independent of the string velocity. This is so, because, although
a denser network will create more wakes per horizon volume the length
of such wakes will be shorter corresponding to filaments of small
linear mass density. Thus, superconducting strings in grand unified
theories (GUTs) could still account for the large scale structure
observed even if they carry substantial currents. 

Similar results are obtained for the temperature anisotropies in the
microwave sky. The later are produced due to the boost of radiation
from the string deficit angle. The anisotropy generated by a single
string is given by,

\begin{equation}
(\frac{\Delta T}{T})_{_{S}}\sim (G\mu)v_{T}\leq G\mu
\end{equation}

However, the overall anisotropy includes contributions of all the
strings that have crossed the line of sight until the present time. An
analytical model to calculate the rms anisotropy from a string network
is described in \cite{leandros}, where it is shown that,
\mbox{$(\Delta T/T)_{rms}\simeq (G\mu)v\sqrt{M}$}, where
\mbox{$M\simeq (H^{-1}/R)^{2}$} is the number of open strings inside a
horizon volume, with $H^{-1}$ being the Hubble radius. Thus, the rms
temperature anisotropy in the case of a plasma scaling string network
is, 

\begin{equation}
(\frac{\Delta T}{T})_{rms}\simeq (G\mu)v_{T}\,\frac{H^{-1}}{R_{pl}}\sim
G\mu
\end{equation}

The above shows that the rms anisotropy is again independent of the
string velocity. Thus, GUT superconducting strings could generate the
observed anisotropy regardless of their current. 
The only effect that a denser network would have on the pattern of
temperature fluctuations would be to hide its non-Gaussian profile in
smaller angular scales than the usually estimated $1^{\circ}$
\cite{eust}, since the inter-string distance would be smaller than the
horizon size at decoupling. 

Superconducting strings could add to the temperature anisotropies by
emitting radiation themselves, mostly due to the decay of loops or
small scale structure. However, as shown in \cite{field}, the
radiational contribution of the open string network is of minor
importance. The later could have an effect regarding only ultra high
energy radiation \cite{schram} and could be a $\gamma$-ray source.

The plasma--scaling solution for the open string network is a direct
consequence of assuming that intercommuting produces loops with
efficiency of order unity. This scaling solution, however, is much
different from the usual horizon scaling of non-superconducting
strings, since the network is denser with slower moving strings. Note
that, in this case, the friction force never becomes negligible, i.e. 
{\em the friction domination era never ends}.

In the above calculations we have implicitly assumed that the
overall energy density is not dominated by the loops produced by the
network. This is not a trivial assumption since current carrying
string loops can avoid total collapse by forming stable vortons which
could have lethal consequences to the universe evolution
\cite{vortons}. However, vorton production is beyond the scope
of this paper. 

\bigskip

{\bf 5. In a primordial magnetic field}

\nopagebreak[4]

\bigskip

\nopagebreak[4]

It would be interesting to embed the whole network in a primordial
magnetic field and observe how its evolution is going to be (if at
all) affected. It is obvious that, in principle, the current carrying
strings do interact with an external magnetic field, since, at a
distance, they appear not too different from current carrying
wires. Thus, the magnetic force per unit length on the stings would
be,

\begin{equation}
f_{B}\sim JB\sin\theta
\label{fB}
\end{equation}
where $B$ is the magnitude of the external field and $\theta$ is the
angle between the magnetic field lines and the string
segment. Depending on $\theta$ the above force can
accelerate or decelerate the string. 

The magnetic force can be compared with $f_{ns}$
and $f_{pl}$ in order to determine under which conditions it will be the
dominating one. For simplicity, we will consider the high temperature
case only.

If $f_{ns}$ is the dominant friction force then by comparing
(\ref{fns}) and (\ref{fB}) (with \mbox{$\sin\theta=1$}) it can
be found that the magnetic force will dominate at,

\begin{equation}
t^{B}_{c}\sim\frac{\eta^{12}}{(B_{0}J)^{4}m_{P}}
\label{tBc}
\end{equation}
where $B_{0}$ is the magnitude of the magnetic field at network
formation. 

The magnetic force will never dominate provided the network reaches
horizon--scaling before $t^{B}_{c}$. The condition for this is,

\begin{equation}
B_{0}<(\frac{J_{c}}{J})\,\eta^{2}
\label{nsB}
\end{equation}

If $f_{pl}$ is the dominant friction force, then
by comparing (\ref{fpl}) with (\ref{fB}) we find that the
magnetic force is subdominant if,

\begin{equation}
B_{0}<\sqrt{\frac{J_{c}}{J}}\,\eta^{2}
\label{plB}
\end{equation}

In order for the magnetic field
energy density \mbox{$\rho_{B}\sim B^{2}/8\pi$} not to
dominate the overall energy density of the universe, we require,
\mbox{$B_{0}\leq\eta^{2}$}. In view of (\ref{nsB}) and (\ref{plB})
this constraint implies that, \mbox{$J<J_{c}$} ensures that the magnetic
field will never dominate the forces acting on the strings.
Thus, {\em a weak current \mbox{$J\leq J_{c}$}
suggests that the string network will follow the standard
non-superconducting string network evolution regardless of the
existence of a primordial magnetic field}. 
On the other hand, in the case of a strong current the magnetic field
can influence the network evolution provided it is stronger than the
constraint (\ref{plB}). 

If the magnetic force was dominant then the network would evolve
according to its action on the strings. In this case 
however, the curvature radius would not grow larger than the coherence
scale of the magnetic field $R_{B}$. Indeed, since the string tension
is dominated by the magnetic force, there is no driving force to
``straighten'' the strings over larger scales. The incoherence of the
magnetic field would twist the strings and curve them over scales of
order $R_{B}$. Also, any loops with dimensions larger than $R_{B}$, 
will not contract. So, over these scales, no string length is lost
and, effectively, there is no loop production. 
Thus, \mbox{$R\propto a$} and we have string domination
\cite{kibb2}. Although the above have been calculated in the high
temperature case only we expect similar results in low temperatures. 

The ratio of energy densities is,

\begin{equation}
\frac{\rho_{s}}{\rho}\sim(\frac{\eta\,t}{R_{B}m_{P}})^{2}
\end{equation}

Thus, string domination could be avoided only if,

\begin{equation}
R_{B}>\frac{\eta}{m_{P}}\,R_{H}
\end{equation}
where \mbox{$R_{H}\sim t$} is the horizon size. For current values
\mbox{$R_{B}\sim 1\,kpc$} and \mbox{$R_{H}\sim 10^{4}Mpc$} we find
that we can avoid string domination if \mbox{$\eta<10^{11}GeV$}.

The above is dependent on the assumption that the string current $J$
remains constant. However, if the external magnetic field is strong
enough, it would affect the string current.\footnote{One could argue
that, since the current grows as \mbox{$\dot{J}\sim vB$} \cite{witt}, 
it will eventually reach its maximum value,
\mbox{$J_{max}\sim\eta$} and from then on remain constant.} Still,
even by taking into account the back-reaction of the external field on
the string current, the qualitative picture is not significantly
changed. 

\bigskip

{\bf 6. The string current assumptions}

\nopagebreak[4]

\bigskip

\nopagebreak[4]

In the above we have assumed that the current switches on at the time
of formation of the string network and remains constant during the
subsequent network evolution. In this paragraph we take a closer look
at these assumptions.

\bigskip

{\em i) Current conservation}

\nopagebreak[4]

\bigskip

\nopagebreak[4]

Witten has suggested that the superconducting strings will most
probably carry a strong DC current, which would persist due
to topological index theorems \cite{witt}. In general, current
conservation is a direct outcome of the field equations for a straight
and infinite superconducting string (see for example \cite{cons}).
However, for a realistic string there are some delicate points
that have to be clarified for our assumption to be justified.

One point is the fact that Brownian contraction gradually decreases
the string length $L$ between two distant fixed points on the string. 
Since, \mbox{$L\sim d^{2}/R$} \cite{book} where $d$ is the distance
between the points, this process suggests that,
\mbox{$\dot{J}/J\sim\dot{R}/R$}. However, this effect is counteracted
by the smoothing of the current flow.

The string current has its own correlation length
$l$. The orientation of the current on larger scales
follows a one-dimensional random walk pattern. Thus, between
two points with string length distance \mbox{$L>l$}, the average
string current is \mbox{$J_{rms}\sim J/\sqrt{(L/l)}$}, where
$J$ is the coherent current inside a correlated string segment. 
Current conservation, suggests that the overall
current $J_{rms}$ remains constant if the string length is unaltered. 
Therefore, the local current has to diminish with time as $l$
grows.\footnote{The growth of the current coherence length is due to
the algebraic addition of the current at the interface between two
initially uncorrelated domains, as have been shown numerically in
\cite{matz}.} 

Putting together the two effects described above we see that the
Brownian contraction would in fact increase $J_{rms}$
with time and this may be enough to hold the local current $J$ more or
less constant, although its coherence length would grow. Note that in
our treatment of the string system we were dealing, in fact with the
local, coherent current on the string. The balance between the two
effects depends delicately on the growth rate of the current coherence
length, which is as yet unknown. A reasonable guess would be that
that the correlation length grows with the speed of light, since,
inside the string, the charge carriers are massless. 

Balancing the above effects suggests that,
\mbox{$J_{rms}L\sim\sqrt{Lt}\,J$}, where
\mbox{$\dot{J}_{rms}/J_{rms}\sim\dot{R}/R$}. Thus, the time dependence
of the current is,

\begin{equation}
\frac{\dot{J}}{J}=\frac{1}{t}-\frac{\dot{R}}{R}
\label{dotJ}
\end{equation}

The above suggests that the current is indeed constant at a scaling
solution, when \mbox{$R\propto t$}. Thus, although the current will be
time dependent during $f_{ns}$ domination our results
are unaffected since, in this period, the existence of a current
does not influence the evolution of the network. 

By means of (\ref{dotJ}), the initial critical current for the
network can be found with the use of (\ref{t*}) and (\ref{Jc}) as,

\begin{equation}
J_{c}^{0}\sim J_{c}\sqrt{\frac{m_{P}}{\eta}}
\end{equation}
for both low and high temperatures.

However, apart from the above, the current built-up inside the strings
could be affected by a number of other effects. Most of these effects
concern the small scale structure of the strings, which may introduce
an AC component to the string current
\cite{nat2}\cite{aryal}\cite{gibb}. However, the AC currents are
likely to be suppressed by radiational back-reaction \cite{evol} and
particle production in cusps \cite{ngrays}\cite{amst1}.

The evolution of the current magnitude on the open
strings is still unclear. If the current was not kept constant,
then its variation could destabilise
the delicate balance of the plasma--scaling solution.
In the case that the string current decreased with time,
$f_{pl}$ would become less effective, the
curvature radius would grow faster than in (\ref{Rpl}) and, thus, it
would eventually catch up with the horizon resulting into horizon
scaling. If the opposite was true and the current
increased with time, then plasma friction would become larger and the
curvature radius would not be able to follow the growth of the horizon
at a constant ratio. Instead the strings would become slower and the
network denser. However, the growth of $J$ would have to end when it
reached $J_{max}$. Then, the network would assume the plasma--scaling
solution at its maximum density case. In all cases, though, we 
still manage to avoid domination of the universe energy density from
the open string network.

\bigskip

{\em ii) Initiation of the current}

\nopagebreak[4]

\bigskip

\nopagebreak[4]

Although we have assumed that the current switches on at network
formation, in general, this could occur at a later stage.
However, even if the
network evolves initially without the presence of a current, the final
picture is not severely altered.  

Indeed, were this the case, the network would initially evolve
according to the standard cosmic string evolution scenario.
Thus, the curvature radius would grow as in (\ref{Rns1})
until it reached horizon scaling. If the time $t_{s}$, when
current switched on, was later than this then the network evolution
would not be affected. If, on the other hand, it switched on before
$t_{*}$ then the evolution would be identical to the one described
earlier provided \mbox{$t_{s}\leq t_{c}$}. If, however,
\mbox{$t_{*}>t_{s}>t_{c}$},  then the string network would remain
comovingly frozen due to excessive plasma friction, until 
the length-scale given by (\ref{Rpl}) reached the size of the 
network curvature radius.
From then on the evolution would continue again as
described earlier. 

Switching on the current in later times could have an effect on the
magnitude of $J$, since the later cannot be larger than the energy
scale at the time. Indeed, suppose that the current switched on at the
temperature, \mbox{$T_{s}\sim\zeta\eta$}, where \mbox{$\zeta\leq
1$}. Then, over a string segment of dimensions of the order of the
curvature radius $R(t_{s})$, the maximum average current
at the current initiation time $t_{s}$  would be,
\mbox{$(J_{rms})_{max}\sim\zeta\eta/\sqrt{n}$}, where \mbox{$n\sim
R(t_{s})/(\zeta\eta)^{-1}$}. Since $R$, until $t_{s}$, would evolve
according to (\ref{Rns1}) the current would be, 

\begin{equation}
(J_{rms})_{max}\sim\zeta^{7/4}\eta\,(\frac{\eta}{m_{P}})^{1/4}
\label{Jbar}
\end{equation}

Assuming that the coherence length of the current grows with
light-speed, after some time the current in the
segment considered, will become coherent with magnitude
\mbox{$J\simeq J_{rms}$}. Thus, (\ref{Jbar}) will
be the maximum possible string current. Comparing with $J_{c}$ of
(\ref{Jc}) we find that, in order for the current to affect the string
network evolution $\zeta$ has to be greater than,

\begin{equation}
\zeta_{c}\equiv (\frac{\eta}{m_{P}})^{3/7}
\end{equation}

For GUT strings \mbox{$\zeta_{c}\sim
10^{-2}$}. Thus, the evolution of GUT strings that become current
carrying at the electroweak transition \cite{an} (\mbox{$\zeta\sim
10^{-14}$}) cannot be affected by their current.

\bigskip

{\bf 7. Discussion and conclusions}

\nopagebreak[4]

\bigskip

\nopagebreak[4]

In conclusion, we have investigated the evolution of a
charged-current carrying, open string network. We
have shown that, in the absence of a primordial magnetic field,
the network, in general, reaches a scaling solution. This ensures
that {\em the network does not dominate the energy density of the
universe}. If the network is embedded in a strong enough magnetic
field, then it is possible that it will never reach a scaling solution
and will dominate the energy density of the universe.

We have also demonstrated that, in all cases considered, the existence
of a current on the strings will have an effect only if the current
is larger than a critical value $J_{c}$, given in (\ref{Jc}). We have
found a similar critical value for the possible influence of a
primordial magnetic field. It is interesting that $J_{c}$ is also the
critical current with respect to radiation emission from the string, 
over which electromagnetic radiation dominates gravitational radiation
\cite{turok}. 

For string current stronger than $J_{c}$ we have shown that
friction never ends and the scaling solution is very different than
the standard cosmic string horizon--scaling. The curvature radius and
the inter-string distance follow the horizon growth in constant
proportion, but they could be much smaller than 
the horizon size. As a consequence, the string network would be
a lot denser. Inside a horizon volume 
the strings would be more curved and twisted and would move
much slower. Thus, the loops produced by the network are smaller,
although the intercommuting rates are unaffected.
Therefore, even though the network is denser, it loses less mass by
loop production and this is why so much of the string length is kept
in the open strings. We called this scaling solution plasma--scaling.

If the network reached plasma--scaling then there are a number of
consequences that may have observational importance. First of all,
production of smaller loops could relax the vorton constraints
\cite{vortons}. Also, a denser network 
would generate large scale structure with much different features
than the one produced by a horizon scaling network. Indeed, the slow
moving strings would create filaments instead of thin wakes, whose
separating distances could be much smaller than the horizon. Also
the imprint of the strings on the microwave sky would be Gaussian in
smaller angular scales than the horizon scale at decoupling. However,
neither the magnitude of the overall density perturbations or the rms
temperature fluctuations will be affected. Therefore, GUT
superconducting strings can still satisfy observations even if they
carry a substantial current. 

The plasma scaling solution could also have important astrophysical
effects. Indeed, a denser superconducting string network would result
in substantial generation of high energy radiation 
\cite{astro}\cite{schram}\cite{ngrays}\cite{grays}. 
Agreement with observations by adjusting accordingly the parameters of
the model could provide information on the underlying theory. 

The evolution of the open string network described in our paper is
expected to be modified in the matter era due to streaming velocities
developed by the plasma during the gravitational collapse of the 
protogalaxies. Such streaming velocities will tangle the strings
since, if friction is dominant, the later are more or less
``glued'' to the plasma \cite{vil}. The situation resembles the case
of a dominant primordial magnetic field. The network
curvature radius and inter-string distance would follow the scale of
the plasma streaming and this could lead to string domination. 

In our treatment we have made a number of assumptions. Since, in
order to explore the curvature radius evolution, we were primarily
interested on larger scales, we chose to ignore the small scale
structure of the strings and its consequences (AC currents, string
linear energy density and tension renormalisation) since it is likely
to be substantially suppressed by radiation back-reaction and particle
production. We have also assumed that the string magnetocylinder is
impenetratable and free of plasma. It can be shown that 
this assumption is valid for \mbox{$J\geq m_{0}\sim 1\,GeV$}.

One fundamental assumption made concerns
current conservation. We have argued that the local string DC
current would remain more or less constant. 
However, more work is required here, particularly on the
evolution of the current's coherence length and the current's AC
component. However, although a variable current
may influence our results, it would never lead to string domination. 
Finally, we have not considered the spring/vorton problem, which is
beyond the scope of this paper.

This work was partly supported by PPARC,
the E.U. under the HCM program (CHRX-CT94-0423), the Isaac Newton Fund
(Trinity College, Cambridge), and the Greek State Scholarships
Foundation (I.K.Y.). We would like to thank \mbox{B. Carter},
\mbox{P. Peter} and \mbox{N. Turok} for discussions. 
Finally, we thank Observatoire de Paris (Meudon) for hospitality.


\begin{thebibliography}{blabla}

\bibitem{witt}
E. Witten, Nucl. Phys. B {\bf 249} (1985) 557.

\bibitem{astro}
E.M. Chudnovsky and A. Vilenkin, Phys. Rev. Lett. {\bf 61} (1988)
1043; \mbox{R. Plaga} Phys. Rev. D {\bf 47} (1993) 3635.

\bibitem{field} 
E.M. Chudnovsky, G.B. Field, D.N. Spergel and \mbox{A. Vilenkin}, 
Phys. Rev. D {\bf 34} (1986) 944.

\bibitem{schram}
C.T. Hill, D.N. Schramm and T.P. Walker, Phys. Rev. D {\bf 36} (1987)
1007. 

\bibitem{ngrays}
B. Pacy\'{n}ski, Ap. J. {\bf 335} (1988) 525.

\bibitem{grays}
R. Plaga, Ap. J. {\bf 424} (1994) L9.

\bibitem{vortons}
\mbox{R.L. Davis} and \mbox{E.P.S. Shellard},
Nucl. Phys. B {\bf 323} (1989) 209;
\mbox{R.Brandenberger}, \mbox{B.Carter}, \mbox{A.C. Davis} and
\mbox{M. Trodden}, Phys. Rev. D {\bf 54} (1996) 6059; 

\bibitem{evol}
M.N. Butler, R.A. Malaney and M. B. Mijic, Phys. Rev. D {\bf 43}
(1991) 2535.

\bibitem{amst1}
P. Amsterdamski, Phys. Rev. D {\bf 39} (1989) 1524.

\bibitem{matz}
P. Laguna and R.A. Matzner, Phys. Rev. D {\bf 41} (1990) 1751.

\bibitem{nat2}
D.N. Spergel, W.H. Press and R.J. Scherrer, Phys. Rev. D {\bf 39}
(1989) 379; Nature {\bf 334} (1988) 682.

\bibitem{warren}
W.B. Perkins, Nucl. Phys. B {\bf 364} (1991) 451.

\bibitem{ever}
A.E. Everett, Phys. Rev. D {\bf 24} (1981) 858;
\mbox{R. Brandenberger}, \mbox{A.C. Davis} and \mbox{A. Matheson},
Nucl. Phys. B {\bf 307} (1988) 909; \mbox{W. Perkins}, 
\mbox{L. Perivolaropoulos}, \mbox{A.C. Davis}, \mbox{R. Brandenberger}
and \mbox{A. Matheson}, Nucl. Phys. B {\bf 353} (1991) 237.

%\bibitem{ann}
%A.C. Davis and W.B. Perkins, Phys. Lett. B {\bf 393} (1997) 46.

\bibitem{kibb}
T.W.B. Kibble, J. Phys. A {\bf 9} (1976) 1387; 
\mbox{T.W.B. Kibble}, Acta Phys. Pol. B {\bf 13} (1982) 723; 
\mbox{T.W.B. Kibble}, Phys. Rep. {\bf 67} (1980) 183.

\bibitem{hin}
M.B. Hindmarsh and T.W.B. Kibble, Rep. Prog. Phys. {\bf 58} (1995)
477. 

\bibitem{carl}
C.J.A.P. Martins and E.P.S. Shellard, Phys. Rev. D {\bf 54} (1996)
2535. 

\bibitem{EW}
J. Ambj{\o}rn and P. Olesen, Int. J. Mod. Phys. A {\bf 23} (1990)
4525; \mbox{W.B. Perkins} and \mbox{A.C. Davis}, Nucl. Phys. B {\bf
406} (1993) 377.

\bibitem{hurley}
J. Hurley, Phys. Fluids {\bf 4} (1961) 109.

\bibitem{book}
A. Vilenkin and E.P.S. Shellard, {\em Cosmic Strings and other
Topological Defects}, Cambridge Monographs on Mathematical Physics,
CUP, Cambridge 1994.

\bibitem{vach}
T. Vachaspati and A. Vilenkin, Phys. Rev. D {\bf 30} (1984) 2036.

\bibitem{wakes}
J. Silk and A. Vilenkin Phys. Rev. Lett. {\bf 53} (1984) 1700;
\mbox{T. Vachaspati}, Phys. Rev. Lett. {\bf 57} (1986) 1655;
\mbox{T. Hara} and \mbox{S. Miyoshi}, Prog. Theor. Phys. {\bf 77}
(1987) 1152.

\bibitem{leandros}
L. Perivolaropoulos, Phys. Lett. B {\bf 298} (1993) 305.

\bibitem{eust}
B. Allen, R. Caldwell, E.P.S. Shellard, A. Stebbins and
S. Veeraghavan, Phys. Rev. Lett. {\bf 77} (1996) 3061;
\mbox{D. Coulson}, \mbox{P. Ferreira}, \mbox{P. Graham} and
\mbox{N. Turok}, Nature {\bf 368} (1994) 27;
\mbox{L. Perivolaropoulos}, Phys. Rev. D {\bf 48} (1993) 1530;
M.N.R.A.S. {\bf 267} (1993) 529; (1997) CRETE-97-14
(astro-ph/9704011). 

\bibitem{kibb2}
T.W.B. Kibble, Phys. Rev. D {\bf 33} (1986) 328.

\bibitem{cons}
A. Vilenkin and T. Vachaspati, Phys. Rev. Lett. {\bf 58} (1987) 1041. 

\bibitem{aryal}
M. Aryal, A. Vilenkin and T. Vachaspati, Phys. Lett. B {\bf 194}
(1987) 25.

\bibitem{gibb}
D.N. Spergel, T. Piran and J. Goodman, Nucl. Phys. B {\bf 291} (1987)
847. 

\bibitem{an}
A.C. Davis and W.B. Perkins, Phys. Lett. B {\bf 390} (1997) 107.

\bibitem{turok}
E. Copeland, D. Haws, M. Hindmarsh and N. Turok, Nucl. Phys. B
{\bf 306} (1988) 908.

\bibitem{vil}
E. Chudnovsky and A. Vilenkin, Phys. Rev. Lett. {\bf 61} (1988) 1043.

\end{thebibliography}
\end{document}